\def\cm2{cm$^{-2}$}

\def\c2{C~{\sc ii}}
\def\c4{C~{\sc iv}}
\def\fe2{Fe~{\sc ii}}
\def\fe3{Fe~{\sc iii}}
\def\mg1{Mg~{\sc i}}
\def\mg2{Mg~{\sc ii}}
\def\si2{Si~{\sc ii}}
\def\si4{Si~{\sc iv}}
\def\al2{Al~{\sc ii}}
\def\al3{Al~{\sc iii}}
\def\o1{O~{\sc i}}
\def\n1{N~{\sc i}}
\def\h1{H~{\sc i}}

\def\approxlt{\mathrel{\spose{\lower 3pt\hbox{$\sim$}}
        \raise 2.0pt\hbox{$<$}}}
\def\approxgt{\mathrel{\spose{\lower 3pt\hbox{$\sim$}}
        \raise 2.0pt\hbox{$>$}}}

\documentclass[article]{gwp80}
\usepackage{graphicx}
\usepackage{amssymb}
\tabletypesize{\normalsize}  

\shortauthors{Kinemuchi}
\shorttitle{Kepler RR Lyrae Stars }

\begin{document}
\large    
\pagenumbering{arabic}
\setcounter{page}{101}

\title{RR Lyrae Research with the Kepler Mission}

%
%
\author{\noindent Karen Kinemuchi{$^{\rm 1,2}$}
{\it (1)  Kepler Guest Observer Office, NASA-Ames Research Center}
}
%
%
\email{(1) Karen.Kinemuchi@nasa.gov}

\altaffiltext{}{(2) Bay Area Environmental Research Institute}

\begin{abstract}

     The Kepler Mission is a Discovery mission supported by NASA's
     Science Mission Directorate, and its primary aim is to discover
     Earth-like planets in the habitable zone of solar-type stars.  The space telescope was designed with a photometer that monitors the Kepler field in a near continuous manner in order to achieve this goal.  With this mission, the asteroseismology community also benefits from the Kepler data via the abundant time-series photometry.  With a short cadence of 1 minute and long cadence of 30 minute observations, the time coverage for many variable stars is unprecedentedly complete.  The Kepler field also contains the archetype RR Lyr, and the Kepler Asteroseismic Science Consortium (KASC) Working Group for RR Lyrae stars have been working to uncover the mysteries surrounding these stars.  I will provide an overview of the Kepler program in relation to RR Lyrae research.

\end{abstract}

\section{Introduction}
The Kepler project, NASA's 10th Discovery Mission, was launched in
March 2009 with the primary objective of finding Earth-sized planets
in the habitable zone of solar-type stars.  While the exoplanet search
is ongoing, the impact to asteroseismology projects is enormous.  The
high precision time-series photometry provided by Kepler is
unprecedented for many areas of variable star research.  For the
celebration conference for George W. Preston, it is fitting to present
the light curve of the eponym RR Lyr from Kepler.

This proceeding is divided as follows: Section 2 covers the
data acquisition of the Kepler data and processing; Section 3 focuses
on some published results of work done of RR Lyrae variable stars found in
the Kepler field of view as well resources available from the Kepler
Guest Observer Office; and Section 4 describes some NASA programs and
opportunities to analyze Kepler data. 

\section{Kepler Observations}
The Kepler field of view is centered at $\alpha_{2000} = 19:22:40$ and
$\delta_{2000} = +44:30:00$ and is $\sim 13.5$ degrees above the
Galactic plane.  Fortunately, the archetype RR Lyr ($\alpha_{2000} =
19:25:27.91$, $\delta_{2000} = +42:47:03.7$) is within the field of
view.  The Kepler photometer has two observational modes, long and
short cadence.  Long cadence observations are 270 coadded images with
a combined exposure time of 30 minutes.  The short cadence, which are 1
minute observations, is only 9 coadds.  The bandpass of Kepler is wide
in the optical regime (4200-9000 \AA).  This bandpass avoids the
calcium H and K lines in the blue end and the fringing effects for the
red end.  The Kepler magnitude is approximately 0.1 magnitudes off of the
$R$ magnitude for almost all stars (Koch et al. 2010).

\begin{figure}
\begin{center}
\includegraphics[scale=0.8]{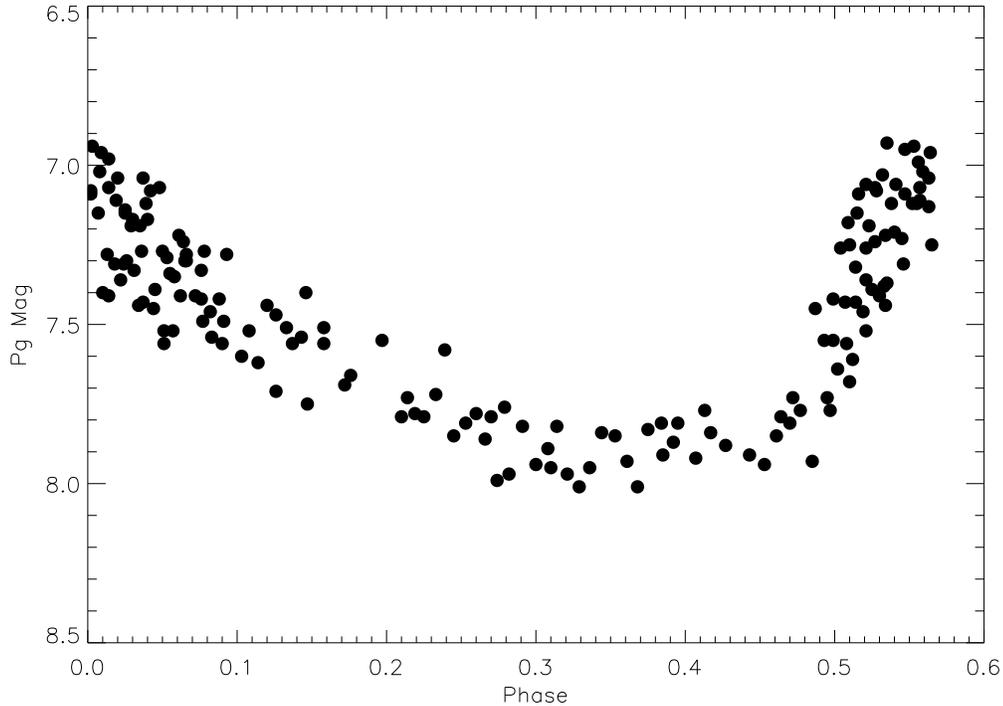}
\caption{Photographic plate measurements of RR Lyr.
  Data taken from table 1 of Tucker (1913). }
\label{photplate}
\end{center}
\end{figure}

Almost 100 years ago, time-series photometry was performed on RR Lyr
with photographic plates. Figure \ref{photplate} shows the light curve of RR Lyr from data reported in
Tucker (1913) that covers 13 days.  The total number of data points is
701 observations to create this light curve.  Even in this data set,
one can see the Blazhko effect modulates the amplitude in
RR Lyr.  The Blazhko effect (Blazhko 1907) has a periodicity between
tens to hundreds of days, and it is characterized by a change in
amplitude and pulsational period. Not all RR Lyrae variable stars exhibit Blazhko, but most
surveys are estimating about 40-50\% of all of these stars have this
effect (Jurcsik et al. 2009, Kolenberg et al. 2010).   Even to this
day, the Blazhko effect still does not have a satisfactory explanation, but with Kepler, we can begin to
discover more about this mysterious phenomenon.  More in-depth
discussions of the Blazhko effect of RR Lyr can be found in this
conference proceeding series by K. Kolenberg or in Kolenberg et al. (2011).

\begin{figure}
\includegraphics[scale=0.5]{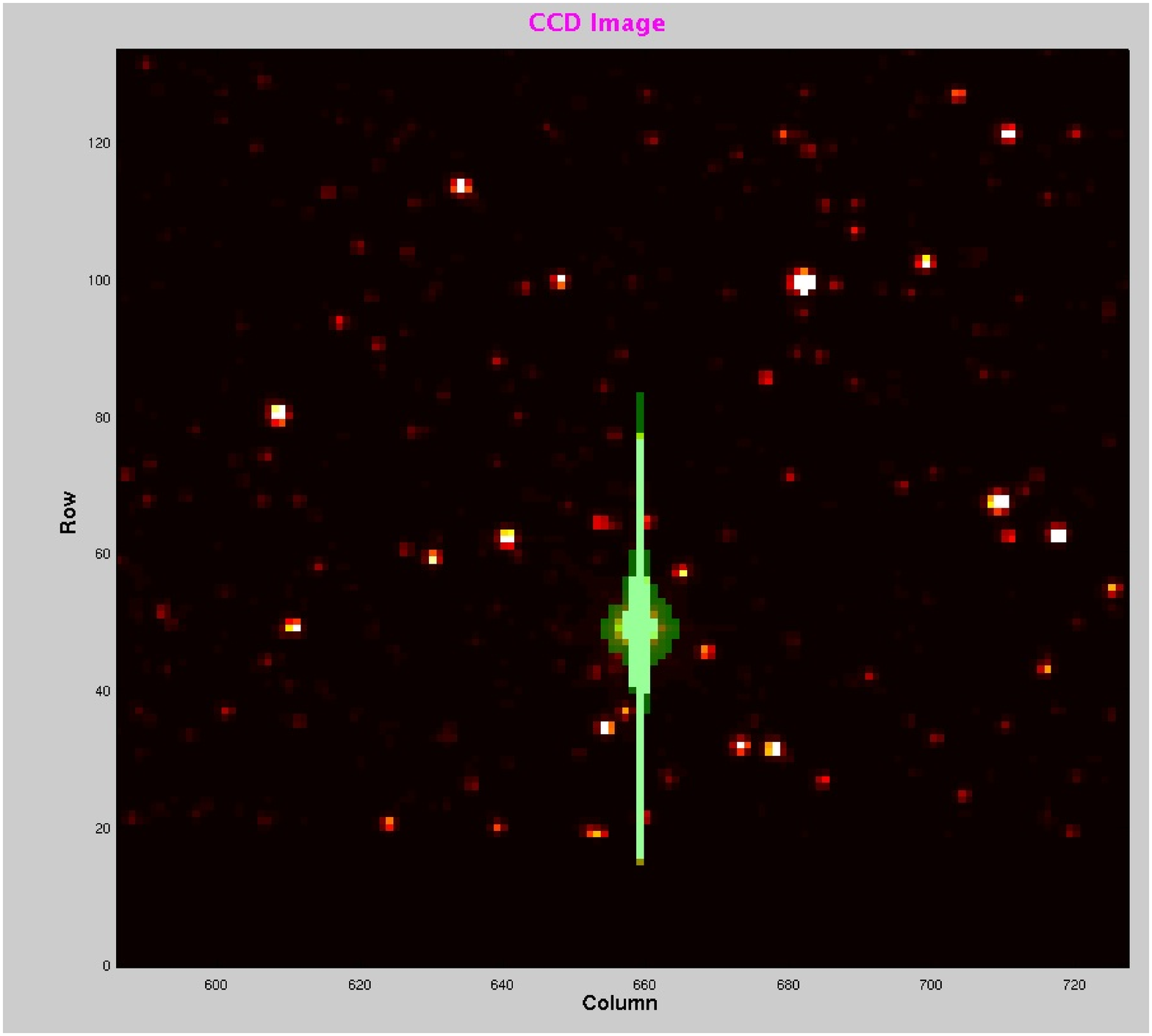}
\caption{Custom optimal aperture for RR Lyr used by Kepler
  spacecraft. The green area is the custom, optimal aperture.  The
  axes mark the location within the CCD chip and are in units of pixels.}
\label{aperture}
\end{figure}

\begin{figure}
\includegraphics[scale=0.8]{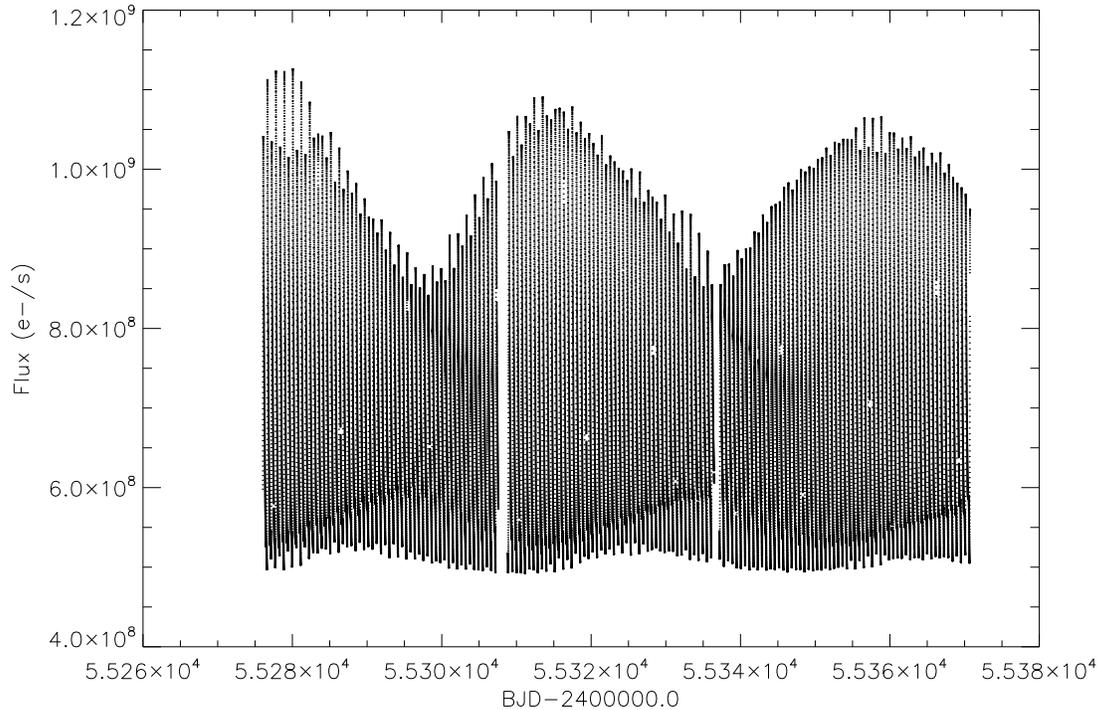}
\caption{Quarter 5 short cadence time-series data of RR Lyr.  This
  data spans March-May 2010.  In this figure, the Blazhko effect of RR
Lyr can be clearly seen.  The Blazhko period is $\sim 39$ days
(Kolenberg et al. 2011).}
\label{quarter}
\end{figure}

\begin{figure}
\includegraphics[scale=0.8]{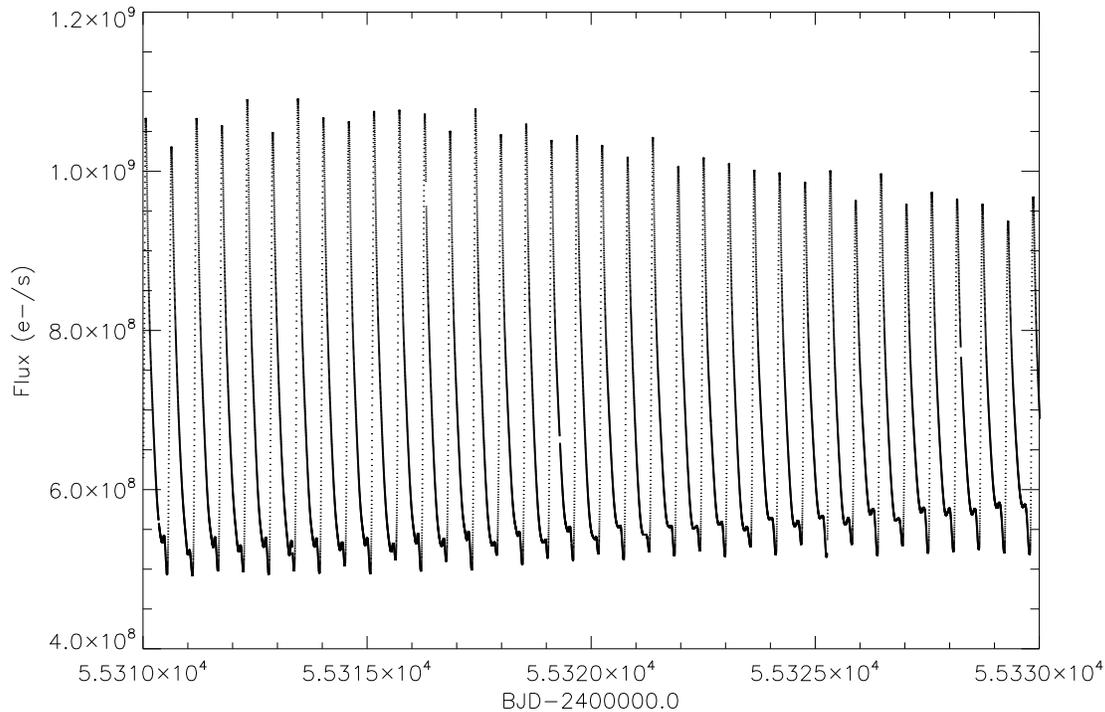}
\caption{Quarter 5 short cadence data of RR Lyr.  Only 1 month of data
are presented here.}
\label{month}
\end{figure}

\begin{figure}
\includegraphics[scale=0.8]{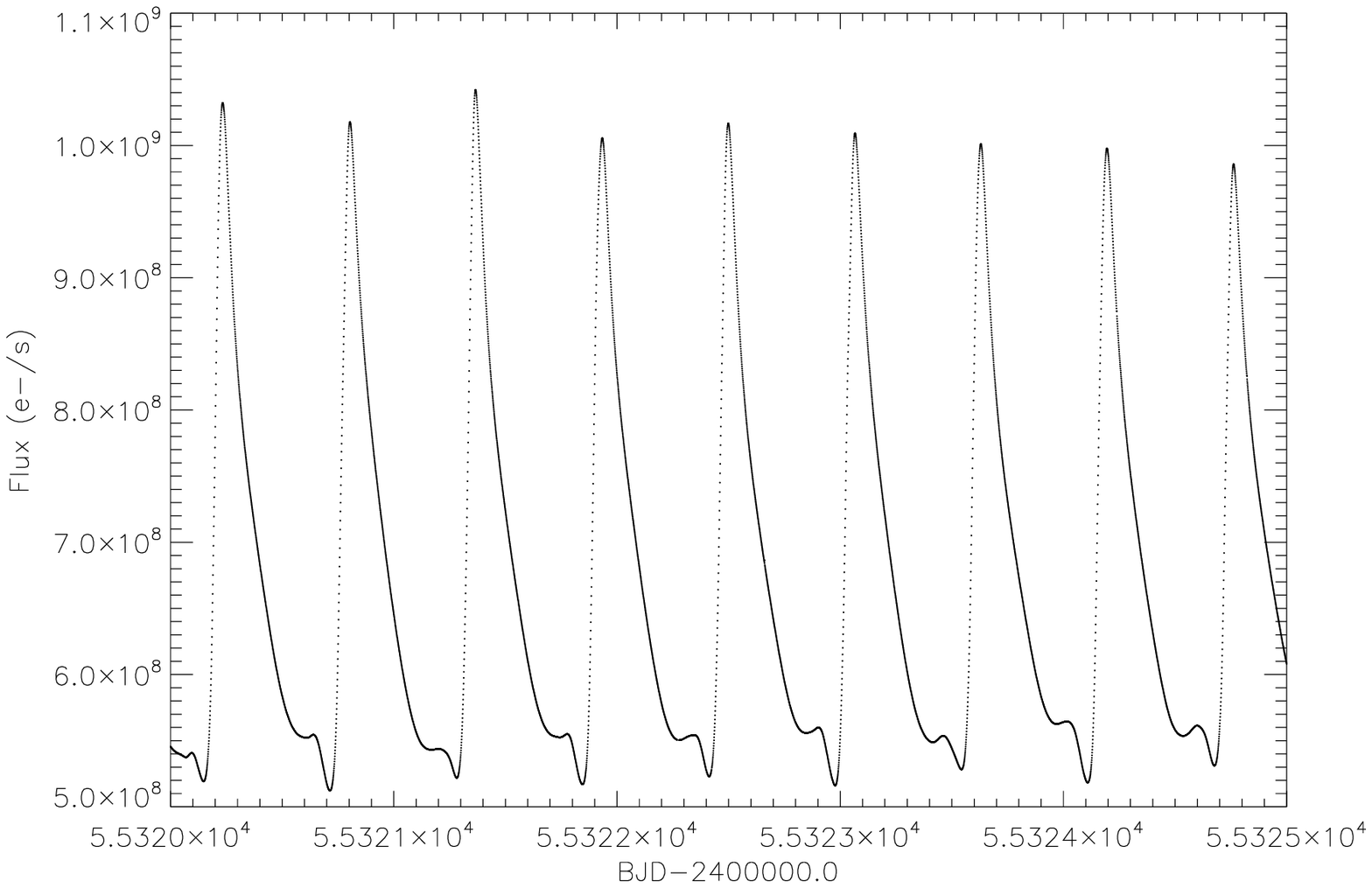}
\caption{Quarter 5 short cadence data of RR Lyr.  A little more than 9
pulsational cycles are seen in 5 days.  A hint of the alternating
maxima can be seen here, which is described as the period doubling
effect (Szabo et al. 2010).}
\label{fivedays}
\end{figure}

Compared to the photographic work done by Tucker (1913), Kepler can
obtain over 100 times more data points with short
cadence mode.  The cadence coverage allows us to observe nearly
continuously the pulsation cycle of RR Lyr, but also many short period variable stars.
Under the Director's Discretionary Target program
\footnote{http://keplergo.arc.nasa.gov/GOprogramDDT.shtml}, short
cadence data were obtained for RR Lyr during Quarter 5 (March-May
2010).  Since RR Lyr is a bright star ($V = 7.1$ magnitude), a special
custom optimal aperture was created.  This optimal aperture takes
into account the bleed columns and essentially recovers all the
photons from RR Lyr.  Figure \ref{aperture} shows the custom aperture
(in green) used for RR Lyr.   In Figure \ref{aperture}, a $10 \times 10$ arcminute region
around the star is shown with the axes in units of pixels.  The pixel
scale of the Kepler photometer is 3.98 arcseconds per pixel.  From the
optimal aperture, the photons were summed from all the pixels assigned
for the aperture.  Even with such a star saturating on the CCD, we are
able to conserve all the flux in the bleed columns in order to measure
the brightness of the star.  With the 1-minute short cadence over
three months, this yielded over 130,000 observations.  The light curve
of RR Lyr is presented in Figures \ref{quarter}, \ref{month}, and \ref{fivedays}.

Due to RR Lyr being such a bright target and the use of a custom
aperture, the data were not processed through the entire Kepler data pipeline,
housed at NASA-Ames Research Center.  Basic data reduction steps were
applied, but the aperture photometry was carefully done outside of the
pipeline in order to produce the light curve information.
Instrumental signatures from the spacecraft are minimal for the light
curve of RR Lyr since it is such a bright star.  However, Kepler users
are advised to consult the Data Release Notes in order to identify any
known instrumental artifacts in the aperture photometry.  For example,
some features include a three-day cycle of the
desaturation of angular momentum of the reaction wheels, initiation of
the long cadence observations, and thermal instabilities, which
affects focus.  

In general, the short cadence data are used for transit timing
measurements for planets, but also for asteroseismology.  In the binary
FITS table files, which can be downloaded from the Multi-Mission
Archive at STScI\footnote{http://archive.stsci.edu/kepler/}
(MAST), there are two data columns used for producing light
curves.  The column labeled ``ap\_raw\_flux''\footnote{ In future releases of the data,
this column of data will be renamed to ``SAP\_FLUX''.} contains the data pipeline
product known as Photometric Analysis or PA.  The data have been
processed with the standard data reductions steps (bias, dark, and
flat fielding) along with Kepler specific calibrations (cosmic ray
removal, gain and nonlinearity, smear, and local detector
undershoot effects).  When this calibrated data goes through the PA
pipeline module, the sky background level is removed and simple
aperture photometry is performed.  The sky background level is
determined from pixels from the long cadence observations and is
extrapolated for short cadence data.  

The column with data labeled as ``ap\_corr\_flux''\footnote{ap\_corr\_flux will be renamed in future
releases of data as ``PDCSAP\_FLUX''} contains data that
has calibrated data (bias, dark, flat fielding, etc has been applied), PA
aperture photometry, and then the Pre-search Data Conditioning  or
``PDC'' processing.  This column contains data where most of instrumental
artifacts have been removed, and the data have been prepared for
planet transit searching.  However, the PDC data or ap\_corr\_flux
values may have much of the astrophysical signatures removed, and the
astronomical community is cautioned in using this data column.   More
details regarding the Kepler pipeline processed data are described in
the Kepler Data Processing Handbook\footnote{can be found at MAST (http://archive.stsci.edu/kepler/) or http://keplergo.arc.nasa.gov/Documentation.shtml}. 

\section{RR Lyrae research}
Currently, there are 37 known RR Lyrae variable stars in the Kepler
field of view, and a majority of them are observed in long cadence.
The Kepler Asteroseismic Science
Consortium\footnote{http://astro.phys.au.dk/KASC/} (KASC) has a
working group devoted to the study of RR Lyrae variable stars.  Recent
results are presented in this conference.  

The first Kepler results of RR Lyrae variable stars are published in
Kolenberg et al. (2010).  As more data are becoming available to the
working group, unique phenomena have been uncovered.  Of the small
sample of RR Lyrae variable stars, 11 exhibit the amplitude and period
modulation known as the Blazhko effect.  Even with the
small sample of RR Lyrae stars, we are finding at least 40\% of the
stars are experiencing the Blazhko effect (Kolenberg et al. 2010).  A few stars with Blazhko also show a secondary,
transient effect previously seen only in RV Tauri type variable
stars.  The period doubling effect (Szabo et al. 2010, Benko et
al. 2010) was observed in 3 Blazhko RR Lyrae stars.  This effect is
characterized by alternating maxima in the Kepler light curves and the
discovery of half-integer frequencies in the frequency spectra.  

To expand the studies of Kepler RR Lyrae stars, we need to find
unknown, fainter candidates.  The confusion limit for Kepler is
approximately $K_{P}=21$ magnitude.  The Kepler Input Catalog (Brown
et al. 2011) contains pre-launch ground based information for objects
down to $K_{P}=17$.  Therefore, there may be a plethora of
unknown variable stars that Kepler could observe. 

A new resource which takes advantage of the monthly full frame images
of the entire Kepler field should list potentially new and unknown variable
stars.  The full frame images are taken by Kepler each month in order
to check the health of the photometer.  These publicly available
images can have aperture photometry performed on them, and thus create
light curves with a much lower cadence.  

To determine which stars are potential variable stars, a difference image was produced.  Those
stars with a large (at least $5\sigma$) residual were considered to be
variable.  Over the entire Kepler field of view, over 260,000 stars
were found to be variable star candidates.  This catalog of variable
star candidates can be used to identify new targets for Kepler to
observe.  For all of these candidates, light curves were produced from Quarter 0
commissioning full frame images.  This set of eight full frame images were
taken when the spacecraft was at optimal pointing and thermally
stable, thus the photometry is not as impacted by many of the instrumental
artifacts which appear in later quarters of data. 

\begin{figure}
\includegraphics[scale=0.5]{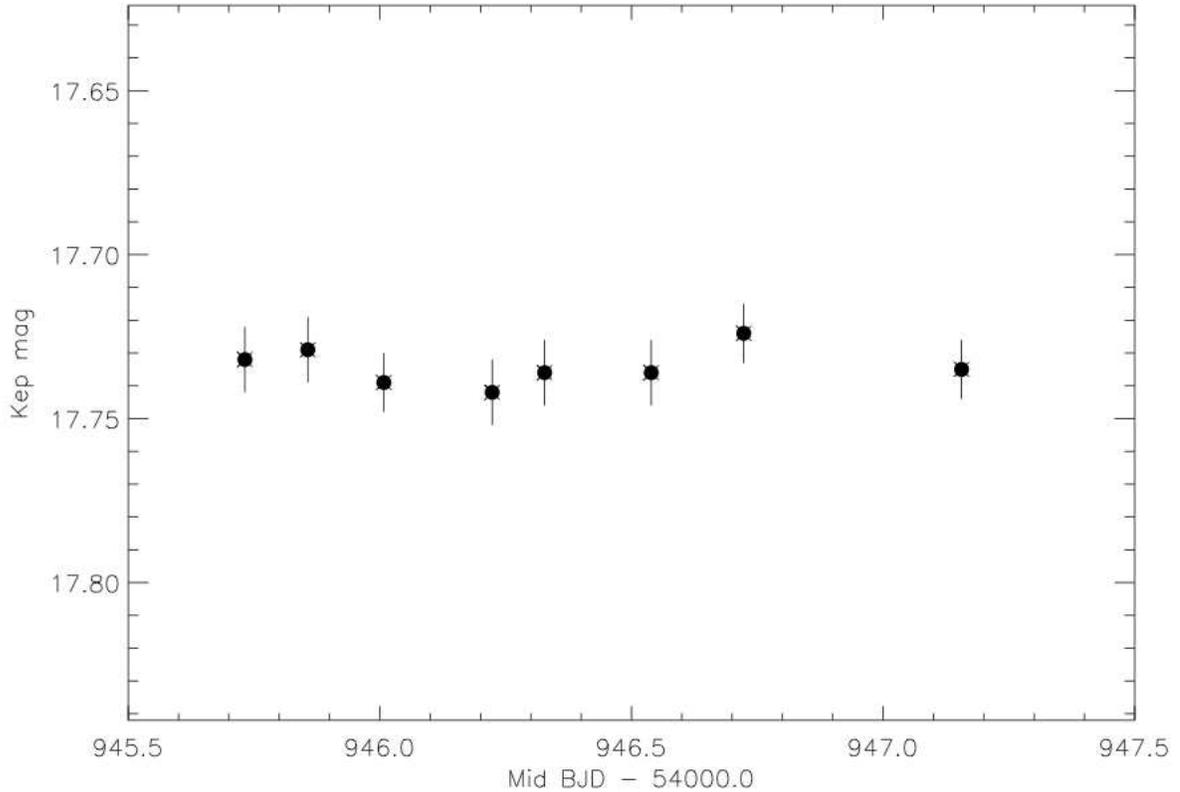}
\caption{Variable star candidate Quarter 0 light curve.  The light
  curve was created from aperture photometry done with the IRAF package apphot on the 8 full
  frame images taken during Kepler's commissioning phase. The y-axis is a relative Kepler magnitude.}
\label{q0ltc}
\end{figure}

\begin{figure}
\begin{center}
\includegraphics[scale=0.35]{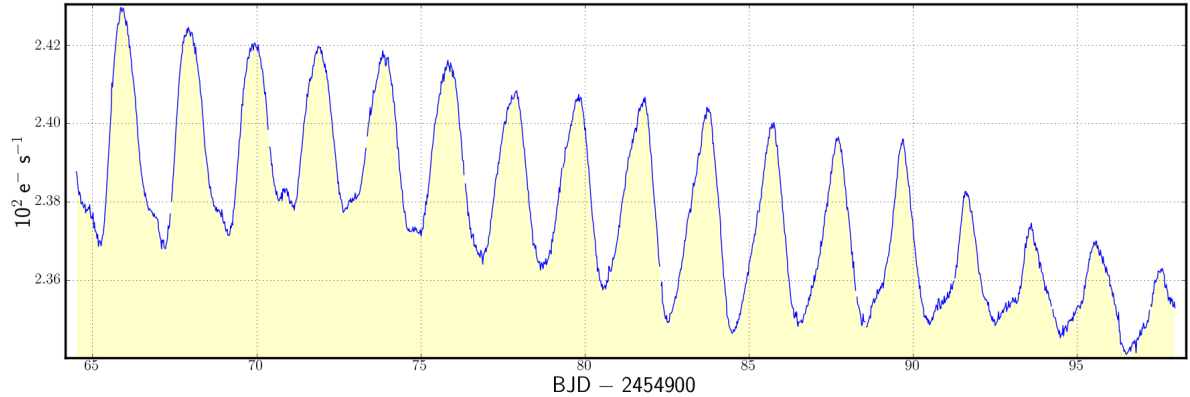}
\caption{Corresponding Quarter 1 light curve for variable candidate in
Figure \ref{q0ltc}.  The long cadence data covers 30 days.}
\label{q1ltc}
\end{center}
\end{figure}

An example of a light curve of a variable star candidate is
shown in Figure \ref{q0ltc}.  These eight data points were not taken
consecutively during the commissioning phase but over a 30 hour
period.  The aperture photometry done on the full frame images was
completed with the IRAF package apphot, thus the uncertainties are due to
the assumption of a Gaussian profile of the star.  The Kepler aperture
photometry uses a more exact pixel response function to maximize the
flux of a target in order to do the photometry (Jenkins et al. 2010).  This star, identified
as Kepler ID 8332007, was also observed during Quarter 1 and with a 30
day coverage in long cadence.  This star appears to be a bona fide variable star as seen in Figure
\ref{q1ltc}.  

Almost 75\% of the variable star candidates that make up this
catalog are not currently being observed by Kepler.  This catalog of
variable stars will provide many areas of asteroseismology more
candidates and better statistics for their variable star studies.
This variable star candidate catalog is slated to be released later in 2011 at the MAST.

\section{Opportunities with Kepler}

The Kepler mission is a resource available to the astronomical
community.  The spacecraft stares nearly continuously at its field of
view, but is travelling in an Earth-trailing orbit.  In the time while
 we can still communicate with the satellite, we need to
take advantage of Kepler's high precision photometry.  As of this
writing, three quarters of data (138 days) are available to the public for both exoplanet and
astrophysical research.  All public data can be retrieved from the
MAST through their various webforms, and the data files are in binary
FITS file format.  

Opportunities exist for the astronomical community to become involved
with Kepler data.   The Guest Observer Program\footnote{http://keplergo.arc.nasa.gov/} will
allow investigators to obtain data on new targets in the Kepler field
of view for astrophysical studies.
The Astrophysics Data Analysis Program supports archival research of
publicly available data.  Both of these NASA programs provide
funding on a yearly basis.  Guest Observer proposal deadlines occur mid-December of each
year of the mission, and the Astrophysics Data Analysis Program
deadlines are mid-May of each year.  Another avenue to obtain data is
through the Kepler Asteroseismic Science
Consortium\footnote{http://astro.phys.au.dk/KASC/}, which is open to
all astronomers. 
On the shorter time scale, Director's Discretionary Target
Program\footnote{http://keplergo.arc.nasa.gov/GOprogramDDT.shtml}
allows the astronomical community to propose new targets to be
observed on a quarterly basis (every 3 months) in either short or long
cadence mode.  This program is excellent for pilot studies in
preparation for either a Guest Observer or Astrophysics Data Analysis
Program proposal.  Deadlines for the Director's Discretionary Target
Program are by the 24th of January, April, July and October.

The main mission statement of the Kepler Guest Observer Office is to
help the astronomical community exploit the resources Kepler provides
through its high precision photometry.  As more quarters of data are
made public, the Office is dedicated in providing support to the
astronomical community through instrument and data expertise.  We
provide access to contributed software
\footnote{http://keplergo.arc.nasa.gov/ContributedSoftware.shtml}
which will help the community analyze the Kepler data.  Not only can RR Lyrae
stars be studied through Kepler, as shown in this proceeding and
others from this conference, but other types of variable stars.
Kepler is expected to expand not only exoplanet investigations but many
areas of astrophysics research.

\newpage

\section{Acknowledgements}
Funding for the Kepler Discovery Mission is provided by NASA's Science
Mission Directorate.  K.K. gratefully thanks G.W. Preston for his
words of wisdom, support, fantastic jokes and stories, and his
kindness.

\end{document}